\documentclass[conference,letterpaper]{IEEEtran}
\usepackage[font=small,skip=10pt]{caption}
\usepackage{enumitem}
\usepackage{tikz}
\usepackage{color}
\usepackage{enumitem}
\usetikzlibrary{calc,trees,positioning,arrows,chains,shapes.geometric, decorations.pathreplacing,decorations.pathmorphing,shapes, matrix,shapes.symbols}
\newcommand{\comment}[1]{}

\tikzset{
>=stealth',
  punktchain/.style={
    rectangle, 
    rounded corners, 
    draw=black, very thick,
    text width=10em, 
    minimum height=2.5em, 
    text centered, 
    on chain},
  punktchain2/.style={
    rectangle, 
    rounded corners, 
    draw=none, very thick,
    text width=10em, 
    minimum height=2.5em, 
    text centered, 
    on chain},
  line/.style={draw, thick, <-},
  element/.style={
    tape,
    top color=white,
    bottom color=blue!50!black!60!,
    minimum width=8em,
    draw=blue!40!black!90, very thick,
    text width=10em, 
    minimum height=3.5em, 
    text centered, 
    on chain},
  every join/.style={->, thick,shorten >=1pt},
  decoration={brace},
  tuborg/.style={decorate},
  tubnode/.style={midway, right=2pt},
}

\begin{document}

%
\title{Cyber Attack Thread: A Control-flow Based Approach to Deconstruct and Mitigate Cyber Threats}

\author{\IEEEauthorblockN{Koustav Sadhukhan\IEEEauthorrefmark{1}, Rao Arvind Mallari\IEEEauthorrefmark{2}}
\IEEEauthorblockA{Scientist, Defence Research and\\Development Organisation, INDIA\\Email: \IEEEauthorrefmark{1}koustavsadhukhan@hqr.drdo.in, \IEEEauthorrefmark{2}arvindrao@hqr.drdo.in}

\and
\IEEEauthorblockN{Tarun Yadav}
\IEEEauthorblockA{Scientist, Scientific Analysis Group\\Defence R \& D Organisation, INDIA\\Email: tarunyadav@sag.drdo.in}}

\maketitle

\begin{abstract}
Attacks in cyberspace have got attention due to risk at privacy, breach of trust and financial losses for individuals as well as organizations. In recent years, these attacks have become more complex to analyze technically, as well as to detect and prevent from accessing confidential data. Although there are many methodologies and mechanisms which have been suggested for cyber-attack detection and prevention, but not from the perspective of an attacker. This paper presents the cyber-defence as hindrances, faced by the attacker, by understanding attack thread and  defence possibilities with existing security mechanisms. Seven phases of \textit{Cyber Attack Thread} are introduced and technical aspects are discussed with reference to APT attacks. The paper aims for security practitioner and administrators as well as for the general audience to understand the attack scenario and defensive security measures.  
\end{abstract}

\IEEEpeerreviewmaketitle

\section{Introduction}

Cyberspace is the most dominant form of communication and data sharing. The privacy of communication depends not only on the trust between two parties, but also depends on trust in the underlying cyberspace where the information is communicated. To maintain this trust information security practitioners implement various security measures on the infrastructure involved. This infrastructure includes various kind of digital systems and networking equipment which are designed with security parameters in mind.
A majority of information sharing and communication happening over this infrastructure is confidential, there are attackers, who are interested and skilled to breach the vulnerable systems to ex-filtrate confidential data.

This paper traverses the attack as it is unleashed by the attacker and tries to explore the path followed by the attacker in context with \textit{Cyber Attack Thread}, which is a derived version of \textit{Cyber Kill Chain}. The paper discusses hindrances faced and methodologies used by the attacker to successfully reach the other end of attack thread. The paper also suggests countermeasures and security mechanisms that a security administrator can deploy to counter the threat from each phase of the attack thread. Suggested Security mechanisms are in a generic form and hence it can be scaled to most platforms of interest.

The paper is organized into 4 sections. Section II defines the term Cyber Attack Thread and discusses its need over Cyber Kill Chain, Section III discusses the technical aspects of Cyber Attack Thread and corresponding defences with available security mechanisms. The paper ends with concluding remarks in Section IV.

\section{Cyber Attack Thread vs. Cyber Kill Chain}

\begin{figure}

\begin{tikzpicture}
  [node distance=.8cm,
  start chain=going below,]
\node[punktchain, join] (reconn) {Reconnaissance};
\begin{scope}[start branch=branch1,]
	\node[punktchain] (AT1) {Attack Vector Verification and Reconnaissance};
\end{scope}

\node[punktchain, join] (weapon)      {Weaponize};

\begin{scope}[start branch=branch2,]
      \node (AT2) [punktchain, on chain=going right] {Enticing the End-User};
      \path [dashed,thick,<-] (AT2) edge node {} (AT1);
 \end{scope}

\node[punktchain, join] (del)      {Delivery};
\begin{scope}[start branch=branch3,]
      \node (AT3) [punktchain, on chain=going right] {Security Penetration to Reach the Target System};
      \path [line,thick] (AT3) edge node {} (AT2);
\end{scope}

\node[punktchain, join] (exploit) {Exploitation};
\begin{scope}[start branch=branch4,]
      \node (AT4) [punktchain, on chain=going right] {Execution on Target System };
      \path [line,thick] (AT4) edge node {} (AT3);
\end{scope}

\node[punktchain, join] (install) {Installation};
\begin{scope}[start branch=branch5,]
      \node (AT5) [punktchain, on chain=going right] { Internal Reconnaissance};
      \path [line,thick] (AT5) edge node {} (AT4);
    \end{scope}

\node[punktchain, join] (cnc) {Command and Control};
\begin{scope}[start branch=branch6,]
      \node (AT6) [punktchain, on chain=going right] {Information Ex-filtration};
      \path [line,thick] (AT6) edge node {} (AT5);
    \end{scope}
    
 \node [punktchain,join ](objectives)   {Act on Objective};
\begin{scope}[start branch=branch7,]
      \node (AT7) [punktchain, on chain=going right] {Covering Tracks};
      \path [dashed,thick,<-] (AT7) edge node {} (AT6);
    \end{scope}
\path [line,thin] (AT1) edge [bend right=10] node {} (reconn);
\path [line,thin] (AT2) edge [bend right=10] node {} (reconn);
\path [line,thin] (AT1) edge [bend right=10] node {} (weapon);
\path [line,thin] (AT2) edge [bend right=10] node {} (del);
\path [line,thin] (AT4) edge [bend right=10] node {} (exploit);
\path [line,thin] (AT4) edge [bend right=10] node {} (install);
\path [line,thin] (AT6) edge [bend right=20] node {} (cnc);
\path [line,thin] (AT5) edge  node {} (objectives);

  \end{tikzpicture}

 \caption{A Shift from Cyber Kill Chain to Cyber Attack Thread} \label{fig:1}
\vspace{-.6cm}
\end{figure}
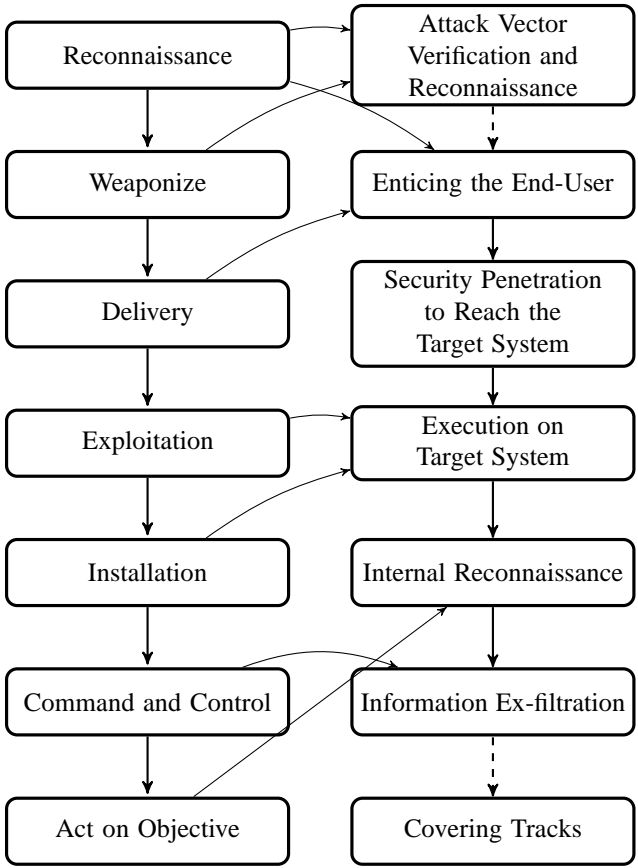

Cyber Kill Chain \cite{sect2:1} is a process based model used by cybersecurity analysts to analyze APT attacks in a chained manner. Such a model enables the analysts to tackle smaller and easier problems, it also helps the defenders to design and develop mitigations and defences for each phase. It describes the 7 stages of an attack with various components \cite{sect2:3} \cite{sect2:4} as shown in Fig. \ref{fig:1}. Understanding Cyber Kill Chain helps a security practitioner to develop defensive counter measures \cite{sect2:2} to the actions of a cyber attacker.

\comment{Details of Cyber Kill Chain with its technical aspects \cite{sect2:5} are discussed previously which .In recent years, cyber-attacks have been more complex and hence more destructive and dangerous. Multiple redundant attack vectors are being used in cyber-attacks to not only multiply the effect but also making it more difficult for the security researchers to analyze. }

There are many articles \cite{sect2:1}\cite{sect2:6}\cite{sect2:7} which describe the cyber kill chain in detail, but are unable to provide a common process model which has the capability to map attack methodology as well as security defences at each stage of the cyber kill chain. Although in \cite{sect2:5} authors discuss technical components and tools involved in Cyber Kill Chain, but it does not help much for a security analyst  to understand that at which stage attack could be prevented and detected in an infrastructure set-up. So there is need for a more generic process model which not only incorporates Cyber Kill Chain, but also helps defender by discussing attacks from his perspective as well.

Cyber Kill Chain explains attacker's strategy\cite{sect2:7} and components  for an attack, but it is not capable enough to map corresponding defence mechanisms in a convenient way. To implement effective securities in the cyberspace it is better to understand the control flow of cyber-attack rather than just its components. To explore the attack there is a need to traverse along with the attack stages and understand evasion techniques for the security mechanisms and impediments faced by the attack during this process. Once the defender knows about the existing security holes which facilitates attacks, and security measures which prevent attacks, it is easier to strengthen defences implemented by him for the infrastructure.

We define control-flow of a cyber-attack as \textbf{\textit{Cyber Attack Thread}}. Cyber Attack Thread gives a simple but insightful methodology to explore the strategy, objectives and technicality of the attack. In this paper a parallel walk with attack thread is executed to analyze how the attack enters, executes, affects and leaves the target with desired information. Security implemented by defenders at network and system layers and hindrances created to attacks are also discussed in detail in the next section.

We suggest 7 phases of \textit{Cyber Attack Thread} as shown in fig \ref{fig:1}. The first phase which is attack design and the last phase covering tracks are oriented \comment{aligned is not a proper word here} towards the attacker rather than the defender. 
Therefore transition from and to these phases are shown as dashed lines. The remaining 5 phases of Cyber Attack Thread are dominant stages of attack where the defender tries to prevent or detect the attack and the attacker tries to evade securities implemented. A mapping of Cyber Kill Chain to Cyber Attack Thread is also shown in fig \ref{fig:1}, which clearly presents control-flow of an attack, which we have defined as Cyber Attack Thread, is different and partially derived from Cyber Kill Chain. In the next section, technical details of Cyber Attack Thread and existing defence mechanisms to prevent and detect attacks are discussed for each of the 7 phases.  

\section{Technical Aspects of Cyber Attack Thread and Corresponding Defences  }
In the previous section we have discussed about the term \textit{Cyber Attack Thread}. Furthermore, there is a need to correlate security measures with attack thread phases for better understanding. In this section technical aspects of attack strategy and existing defence measures are discussed. Attack thread is explained as a 7 stage process in which first and last stages are not much critical from defender's perspective, but are included and discussed for completeness. 

\subsection{\textbf{Attack Vector Verification and Reconnaissance}}
A multi-staged cyber-attack has many components like exploit, RAT, rootkits, etc., which are designed and developed by the attacker to achieve a malicious end result. Before carrying out the actual cyber-attack on the intended target, the attacker integrates all the components and tests it according to a specific target scenario. \textit{Target Scenario} is defined by the hardware and software configuration of the target.

For targeted cyber-attacks, attackers develop specific components according to specific target scenarios and configurations. A typical scenario includes network components, Operating System major and minor versions, software application versions, anti-viruses and other security mechanisms. An attacker does the initial groundwork by testing all the attack components in an emulated environment similar to the target scenario. The attacker also verifies the command \& control communication, data ex-filtration, etc. and minimizes all attribution proofs in the emulated environment. The \textit{target scenario} is created, using the information gathered by carrying out reconnaissance against the target infrastructure. There are basically two types of reconnaissance:
\begin{enumerate}
\item{\textit{\textbf{Passive Reconnaissance:}} Gathering information about the target, without letting the target know about it, is called Passive Reconnaissance. Attackers can gain useful information from the Internet especially social-media about an organization and its employees. This information is used by the attackers to determine the management hierarchy and targets of interest, which is generally used by attackers to plan subsequent attacks. 

\hspace{1em} One approach to minimize the damage from such information is to do an internal assessment. The task of internal team is to act like an adversary to identify valuable information from internet and social media, and then estimate the damage which can be caused by using the information. Another approach can be to create and spread information about fake personas to deceive the attackers about an organization and its employees. This approach will make the attack more costly by misleading the attackers in the wrong direction. 
\cite{phase1:1}} 

\item{\textit{\textbf{Active Reconnaissance:}} When attackers directly engage with the target and scan its network to determine open ports, running services, network security posture etc. to make a list of likely attack vectors, it is known as active reconnaissance. It is deeper profiling of the target which might trigger an alert to the target. The usual method of preventing active reconnaissance is to place a firewall between the internal and external network. The services are configured so that they do not give complete information about version, type and system configuration. \\
\hspace{1em}Active reconnaissance can also be detected by placing a passive network sniffing device between the firewall and the external network. This passive device will log all the traffic coming in and out of the network and can detect any attempt of active reconnaissance. 
\cite{phase1:2}.}
\end{enumerate}



\subsection{\textbf{Enticing the End-User}}
After Phase A, an attacker has sufficient knowledge about the target's affinity and interests. Equipped with this knowledge the attacker will now initiate penetrating target's periphery by enticing him/her. In an attack scenario, periphery is the environment surrounding the target system which is a combination of network and network devices, system software and human interface. 
Entering into the target's periphery depends on the method of delivery of attack. Any affinity shown by the user towards the attack delivery medium is less prone to detection, because of the user's trust in the content. Attacks are targeted on an organization, particular user or the masses. In case of organizations, common interest of content to the employees is used to increase the affinity towards the attack's delivery.  For attacks targeting a particular user, an attacker uses the personal information harvested from stage A. There are two ways how an attacker enters into the target's virtual periphery:
\begin{enumerate} 

\item{\textit{\textbf{Targeting User:}} This mode of operation directly interacts with the user and tries to convince him/her to do the desired activity, widely known as \textit{Social Engineering}. 
In almost all cases, some kind of user interaction is needed to deliver the attack's exploit and payload. This interaction is performed by various kind of methods \cite{sect2:5} mainly phishing, emails, drive-by-downloads, USB/Removal Media and DNS Cache Poisoning.

 \hspace{1em}There are various techniques like spam filers, web-filtering software, heuristic analysis by mail servers and firewall, Google or Bing webmasters, DNSSEC \cite{phase2:1} which give protection against such delivery methods, but there are several ways to bypass such protections, most of the times due to inefficient implementation of these techniques. Other than these techniques, the attacker faces hindrances due to strict policies implemented by security administration which restricts user to use removable media, public domain websites and gives limited access. If an attack passes through the defences explained, then the best protection is user awareness. A user, who is well aware about the common security risks is less likely to be fooled or enticed by any of the mentioned attacker's mode of delivery.
Attackers use active and passive reconnaissance to gather enough information about the target and use this information to serve an appealing content via one of the available delivery modes. Reconnaissance enables the attacker to craft contents according to user affinity, and maximize the necessity of user interaction, which leads to breach into the system or network periphery.
}

\item{\textit{\textbf{Targeting Third-Party Interactions with User:}}  Targeting third-party is an easier and reliable option for an attacker. These third-parties are mainly online websites, organizational/internal portals and network services. As soon as the attacker gets control of these third parties, it alters the content to be served. These altered contents are exploits and payloads. In this case, if any user uses services provided by these third-parties, he will be served with the attacker's crafted content which has the ability to compromise a vulnerable target system. For an attacker it is a better option, but due to segregation of internal portals from the Internet, strict firewall policies for network services, continuous updates and upgrade policies it is not an easy task to find a viable and exploitable third party. As we can clearly understand that in this case there is no direct interaction with targeted user so user awareness at this stage doesn't apply directly, but an aware user is likely to detect abnormal behavior in the system to trigger the alarm.}
\end{enumerate}

\subsection{\textbf{Security Penetration to Reach the Target System}}
As soon as the attacker gains access to the target's periphery, he tries to compromise the target, but many obstacles exist in this process. The attacker has to bypass a variety of security mechanisms implemented by security professionals. Defences in this phase can be primarily categorized into 3 parts: 
\subsubsection{\textbf{Inbound Network Security}}
The first step of an attacker is to penetrate the network perimeter of the target. There are many systems available to protect the network perimeter, which assure high reliability, if implemented and configured properly. Some of the systems are being discussed here.
\begin{enumerate} [label=\alph*)]
\item{\textbf{Intrusion Detection and Prevention Systems:}}
An intrusion detection and prevention system can be software only or mixture of dedicated hardware and software, which is used to monitor suspicious events in end-points or networks. Generally IDPS has the capability to detect and prevent reconnaissance, exploitation and exfiltration events \cite{koustav:1}.

IDS is responsible for automating the process of detection of intrusion. IPS includes all the functionality of IDS and it also contains features to stop and prevent malicious events. Most of the IDPS systems employ the following techniques to detect and prevent malicious activities:
\begin{itemize}
\item{\textit{Signature Based} - This generally relies on pattern recognition techniques, which includes comparison of observed events with the database of signatures for known threats. This approach often proves to be very effective for detecting known threats, but mostly ineffective for unknown threats and zero-days. \cite{koustav:2}\cite{koustav:3}}

\item{\textit{Anomaly Based} -  Anomaly based IDS is based on underlying principle that the normal behaviour of the network traffic is different from that of the behavior of the attack traffic. The major advantage of this approach is that it has the ability to detect previously unknown attacks for which signatures are not available. But on the other side, its disadvantage is that the number of false positives in anomaly based systems is much higher, when compared to the signature based systems. \cite{koustav:4}}

\item{\textit{Stateful protocol Analysis} -  The system first identifies the protocol in use, then it compares the observed events with the stored profiles of benign protocol activity. If it detects any deviations from the normal protocol states, then it raises an alarm. The drawback of the stateful protocol analysis is that it very resource intensive, as it has to continuously keep track of the different states of a protocol for multiple simultaneous connections. Another drawback of this approach is that it cannot detect an attack which does not involve any deviation from normal protocol states. \cite{koustav:5}\cite{koustav:6}\cite{koustav:8}}

\end{itemize}
\setlength{\parindent}{0em}
\textbf{Network Based IDPS Technologies}- SNORT, Suricata and Bro IDS are three major technologies used for network based IDPS. SNORT is the most popular open source network based IPS which can capture network packets as well as perform malicious traffic analysis in real-time\cite{koustav:9}. Suricata is also based on signatures, but it includes some advanced features such as high performance due to multithreading. It also has the ability of automatically recognizing commonly used protocols\cite{koustav:10}\cite{koustav:11}. The main design goal of Bro IDS is to separate policy and mechanism. The working principle of Bro is based on events with custom policies, which is an abstraction of network activity \cite{koustav:12}.



\item{\textbf{Network Based Firewall:}}
It is used to protect the perimeter of a network by monitoring inbound and outbound traffic. Network firewalls protect the perimeter of the network by filtering the traffic based on the criteria and configuration the administrator has defined. Network firewalls also include additional features compare to host-based firewalls like Network adapter input filters, Network address translation (NAT), Static packet filters, Circuit-level inspection,Stateful inspection and  Application layer filtering \cite{phase3:1}.
\end{enumerate}
\subsubsection{\textbf{Inbound System Security}}
After successfully evading network security, the target system is exposed to the attacker's threat. The attacker now drops and tries to execute a malicious payload on the target system, but before executing on the system, its presence can be detected and prevented, using the defensive measures implemented by the target. We categorize these measures into 4 major types: 
\begin{enumerate} [label=\alph*)]
\item{\textbf{Anti-Virus Static Detection:}}
Static detection uses the information of already analyzed malware patterns to categorize, whether a piece of code is malicious or not. These malware patterns are updated by anti-virus engines regularly, to detect a variant or combination of malware.
We categorize static detection technique as follows\cite{phase3:3}:
\begin{itemize}
\item{\textit{Signature Based} -It uses static patterns of an analyzed file to generate fingerprints of known malware component of an attack. The signature could represent a series of bytes or a hash of the file or selected parts. Static detection is an essential component of every anti-virus tool because before execution it alerts for malicious file or program. There are various methods used for static scan e.g. String Scanning method, Wildcards method, Mismatches method,Bookmarks method,Smart Scanning,Skeleton detection. \cite{phase3:2}

Limitation of signature based detection is that it is not able to categorize newly developed malware or malicious files for which signatures have not yet been developed. Therefore, modern attackers mutate attack components often to retain malicious functionality by changing the signatures.}

\item{\textit{Heuristics Based} - Heuristic detection detects new malware and variants of old malware by statically examining files for suspicious characteristics where static detection give no exact signature match. For example, an anti-virus might look for the presence of rare instructions or junk code in the analyzed file. The tool might also emulate running the file to see what it would do if executed, attempting to do this without noticeably slowing down the system. A single suspicious attribute might not be enough to flag the file as malicious. However, several such characteristics might exceed the expected risk threshold, leading the tool to classify the file as malware. The biggest downside of heuristics is the high rate of false positives for legitimate files. }

\end{itemize}


\item{\textbf{Host Based IDS:}}
Like network based IDS, host based IDS tries to detect and protect the host system  from various kinds of malware and malicious activities. The major features of Host based IDPS are file integrity checking, event log analysis, OS configuration monitoring, rootkit detection, port monitoring, detecting hidden processes and generating real-time alerts. OSSEC [25]and Samhain [26] are the two famous host based open source IDS compatible with various operating systems. 

\item{\textbf{Host Based Firewall:}}
 Traditional firewall architectures protect only the network perimeter. Host based firewalls are very effective in preventing attackers  from attacking an individual computer. Generally host based firewall can perform stateful analysis of network traffic and are designed to drop any incoming traffic which does not conform to the rules configured in the firewall. A stateful analysis of network traffic means that firewall keeps track of the state of network connections. The firewall monitors traffic for network connections initiated by the host and allow responses corresponding to the sent traffic by adding exceptions dynamically to its rules. Due to this behaviour, host based firewalls can prevent attackers from attacking computers by blocking incoming malicious network traffic.




  
\item{\textbf{Application Security:}}
 Although the OS implements various security mechanisms to prevent attacks as explained in the previous section, but these days, applications also have their own security mechanisms to prevent malicious files to be downloaded to the system. In most of the browsers, protection against online malicious contents is implemented. Google Chrome, Firefox and Internet Explorer alert users regarding expired certificates, blacklisted websites having phishing pages or malicious contents, unsigned executable and other unsafe downloads. Attacker needs to consider all these detection parameters, while designing and delivering the attack components, mainly the exploit and the payload. Hosting attack components on genuine publicly accessible compromised websites, signing payload with stolen certificates, and use of crypter and packers to change the signature of file are methods, which are used to bypass application security to allow the user to access the content served by the attacker.
\end{enumerate}
\subsubsection{\textbf{User Awareness}} 
User awareness is the security which is not easy to break. Nowadays, cyber security has become an international topic of discussion, so more and more people are well-aware about it, and they keep their system updated and upgraded. In such scenario, it is not easy for the attacker to bypass user awareness, although most of the times social engineering, user history, organization objectives and current topics of interest to user helps a lot for the attacker to deceive the user with the content served. A security aware user will detect malicious activity like crashing applications, modified URL of phishing pages and websites with expired certificates easily and they may scan and report such activities either to administrator or online virus scanner services e.g. Virus Total. So, nowadays attackers are becoming more careful and putting more effort to hide the attack from users, because it may lead to destroy the whole attack campaign. 
\setlength{\parindent}{1em} 
\subsection{\textbf{Execution on Target System}}
The attacker now acquires the ability to execute the crafted malicious content on the target system. This implies that the attacker has already bypassed all static detection mechanism as we explained in the previous section. Now execution of attack's initial component is initiated. At this stage we break the cyber-attack components into 2 parts: 

\subsubsection{\textbf{Exploit}} An exploit works as a carrier for the payload. Exploits are developed using vulnerabilities in various types of software applications or hardware devices e.g. Office software, PDF Readers, Web-Browsers, Operating System and routers. Most of the exploits use memory corruption vulnerabilities, but there are protection techniques, which are implemented by operating system to prevent successful exploitation of vulnerabilities:
\begin{enumerate} [label=\alph*)]
\item{\textbf{Mitigations for Stack Based Exploits:}}

\begin{itemize}
\item{\textit{Stack Cookies or Stackguard} - The stack cookie (as known in Windows) or stackguard (as known in Linux) is a random value which is placed between the stack local variables and stack meta data such as return address. Whenever a function returns, the program verifies the integrity of the stack cookie value. In case, if the value of the stack cookie fails to match the program determines that a stack based buffer-overflow has happened and terminates itself. \cite{koustav:13}}

\item{\textit{Structured Exception Handler Overwrite Protection (SEHOP) }- When a user-mode thread begins its execution, a symbolic exception registration record is inserted as the last entry in the exception handler list. The subsequent exception registration records are appended to the head of the list, thus the symbolic record always remains the last entry of the list. Whenever an exception dispatcher is notified that an exception has occurred in user mode, it walks the exception handler list to check whether the symbolic record can be reached and is valid. If the exception dispatcher is unable to reach the symbolic record, then it deduces that the exception handler list is corrupt. The process is then safely terminated by the exception dispatcher.}
\end{itemize}
\item{\textbf{Mitigations for Heap Based Exploits:}}
\begin{itemize}

\item{\textit{Safe Unlinking} - Whenever a free operation happens in a chunk of heap memory, a verification check occurs, to make sure that the list entry stored in the chunk being freed, is a valid doubly linked list entry. This mitigation prevents exploitation techniques, that rely on the unlinking performed for the coalescing of freed heap chunks to write arbitrary value at an arbitrary location.}

\item{\textit{Allocation Order Randomization} - When exploiting heap corruption bugs, attackers often rely on deterministic heap allocations, so that they can position heap chunks adjacent to each other, or at a desired location. This technique popularly known as heap massaging or heap normalization \cite{koustav:15}, is widely used for exploiting heap-overflow \cite{koustav:16} and use-after-free vulnerabilities \cite{koustav:17}. These kinds of exploitation are mitigated, by  randomizing the order of heap chunk allocations, which means that it is no longer guaranteed, that subsequent heap allocations will be placed next to each other. \cite{koustav:19}}
\item{\textit{Virtual Table Guard} - \textit{Use-after-free} vulnerabilities are generally exploited by creating a fake instance of a C++ object that has a fake virtual function table which contains attacker controlled data. The attacker hijacks the control-flow by forcing a call to the controlled virtual method. \textit{Virtual Table Guard} is a compiler security feature, which places a randomized secret value in the Virtual Function Table. A check is made to verify the integrity of the secret value before a call is made to one of the virtual functions. In case, if the secret value fails to match the application is terminated. \cite{koustav:18}}
\end{itemize}
\item{\textbf{System-wide Exploit Mitigations:}}
\begin{itemize} 
\item{\textit{ASLR} - Address Space Layout Randomization randomizes the base address of the  various segments, such as stack, heap, libraries and the executable itself. Randomized base address of different segments make the location of code and data unpredictable. Code segment layout randomization mitigates code reuse attacks; data randomization hinders control-flow hijacking by making it difficult to predict the location of the attacker injected code. \cite{koustav:20}\cite{koustav:21}}

\item{\textit{DEP or NX} - Data Execution Prevention is a combination of hardware and software which marks all memory locations in a process as non-executable, unless the location is specifically marked for execution. The application terminates as soon as the attacker tries to hijack control by executing from the non-executable pages. \cite{koustav:22}}

\item{\textit{Control Flow Integrity} - Control flow Integrity is a compile and link time exploit mitigation technology, which analyzes a program and discovers every location in the program that can be reached by any indirect branch. This information is then stored as a part of the executable in a separate data structure known as Control Flow Graph. A check is made while the program is running before every indirect branching, to ensure that the target is one of the expected locations. If the program detects an indirect branching to a location, other than the legal targets, it detects an attempt of control-flow hijack, and terminates immediately. \cite{koustav:23}\cite{koustav:24}}
\end{itemize}
\end{enumerate}

\subsubsection{\textbf{Payload}}
Payload is the core component of the cyber-attack which executes on the system. Many kinds of payloads exist e.g. Remote Access Toolkit, rootkit , bootkit, dropper, downloader etc. These payloads optionally have persistent and stealthy installation characteristics to evade detection. Objectives for most of the payloads are to open a backdoor to exfiltrate the data(files,keylogs,user credentials) , gain persistent access and administrator access to modify security configurations, propagation by infecting other systems in network, and sometime attacker aims to damage the system by increasing resource usage to the maximum.  There are various techniques used by the attacker to prevent detection e.g. developing polymorphic and metamorphic payloads, use of crypter and packers, etc.

Anti-Viruses analysis engine is one of the popular security mechanism for a system which tries to prevent and detect malware, during the execution of payload. It basically does 3 types of analysis and then detection \cite{phase3:3} during execution of payload:
 
\begin{enumerate}[label=\alph*)]
\item{\textbf{Heuristic based Analysis and Detection:}
\comment{In previous section we have discussed static heuristic analysis. As explained the static heuristic analyzer is based on signatures and code analysis to categorize the behavior of programs, the } 
A dynamic heuristic scanner performs CPU emulation of malicious code as it happens in an actual system and monitors the control flow to guess the characteristics.

\hspace{1em}Many times malware use crypter and packers as wrappers, which modify the signature as well as initial behavior of the program by means of adding delay or changing the code execution path. The dynamic heuristic engine emulates running the file to analyze the behavior, if executed. Most of the time single suspicious attribute is not be enough to label the file as malicious, therefore multiple attributes are used to check the expected risk threshold, leading to classify the file as a malware. The biggest downside of heuristics is the high number of false positives which flag legitimate files as malicious.}

\item{\textbf{Behavioral Analysis and Detection:} Behavioral Analysis observes the execution of program on the system rather than emulating its execution. This method attempts to detect malware by observing and analysing for suspicious behaviors, such as decryption and unpacking of malicious code, modifying the hosts file properties, and observing uses of system API for keystrokes, file access, network connections, removable media access and modifying process properties, etc. This suspicious behavior is categorized using analysis of the database of detected malware. Monitoring such malicious behavior allows an anti-virus to detect the presence of unknown malware on the protected system. 

\hspace{1em}As happens with heuristics detection, one malicious activity might not be enough to classify the file or code as malware. Whereas collective analysis of such activities could be indicative of the presence of malicious code in a program. Behavioral detection techniques fill the gap between anti-virus tools and host intrusion prevention systems (HIPS), which have traditionally existed as a separate category of security mechanism.
}
\item{\textbf{Cloud based detection:} The process of cloud based detection is based on collecting malware data from protected systems, and analyzing it on the cloud infrastructure, instead of local system where malware resides. The relevant properties of file and context of execution are captured, and submitted to cloud based analysis engine for processing. Then the cloud engine derives patterns related to malware characteristics and behavior, by correlating data from multiple systems. Motivation for cloud-based analysis is to get benefited from the malware experiences of other systems, connected to the cloud, submitting malicious files to the same engine}
\end{enumerate}
Other than anti-viruses, \textbf{\textit{Sandboxing}} is a security technique that isolates programs using a tightly controlled environment, preventing malicious programs from modifying system configuration and data exfiltration. Sandboxes restrict the capabilities of a running program, so that it cannot abuse certain privileged actions on the system. \cite{phase4:1}

Other programs like browsers e.g. Google Chrome and Internet Explorer both run in a sandbox themselves.  of the browser, which deals with untrusted content from the internet, runs at a low or untrusted privilege. So, even if the attacker exploits a vulnerability in the sandboxed process of the browser, he cannot perform any privileged action which can compromise the system, without escaping from the sandbox.

Content like web pages, browser plug-ins, PDFs and Office documents are sandboxed by software applications used to open such content while mobile apps and windows programs are sandboxed using User Access Control mechanism. There are many techniques like virtual machine, sanboxie, AirGap, spoon.net\cite{phase4:2} which are used to implement a sandbox environment.

\subsection{\textbf{Internal Reconnaissance}}
At this phase of the attack thread, the payload developed by the attacker has been delivered and deployed by the exploit. Using one of the persistence mechanisms explained earlier, the payload also has gained partial or complete persistence, depending on the type of privilege escalation achieved. Once the payload has gained a foothold, the first and foremost step it is programmed to do, is to inform the attacker of its success, using the command and control infrastructure set up by the attackers. Based on the way the payload is programmed, it may wait for command from the command and control server, or start collecting confidential information like system configurations, document files, stored passwords etc. For an attacker, there are basically two classes of information that a compromised target can yield:
\begin{enumerate}
\item {\textbf{\textit{Primary Tactical Information:}} Usually, the main intent behind targeted cyber espionage is to compromise the confidentiality of the target. Such attacks aims at smuggling out information regarding the target machine owner’s financial, organizational, academic and/or personal interests. This class of information usually includes confidential documents, presentations, financial records, keylogs, stored credentials, screen shots, stored emails, browsing habits, encryption keys, chat logs, \cite{info:1} etc. Such information gathered are called primary tactical information.}
 
\item{\textbf{\textit{Secondary Strategic Information:}} This is another class of information that can be yielded from a target. Secondary Strategic information is used by the attackers to establish a stronger foothold, and establish a means to spread the payload, to other systems it can connect to. To achieve this objective, the payload surveys the system for the processes running, software installed, services enabled, update history, network configuration, cached DNS servers, etc. The payload can also initiate the installation of more functionalities, like port forwarding, execution of remote exploit, proxy to the internal network and proxy to the World Wide Web.}
\end{enumerate}
To protect sensitive data and reduce the risk of data leakage, organizations use Security Information and Event Management(SIEM) technology. SIEM is equipped with features like Real-Time Event correlation, embedded file integrity monitoring and log/event manager\cite{info:2} to detect and prevent data loss. Organizations also tackle data leak by means of labelling data. Depending on the required confidentiality, policies are in place for each of the label. For example, Top Secret documents should always be password protected while restricted documents don't need to be protected by passwords. As a security practice, storage of credentials should be minimized and two-factor authentication should be used wherever available. Browser Cache should be cleared regularly, and access to system logs should be restricted to specific applications. 
\subsection{\textbf{Information Ex-filtration}}
The attacker by now is using the command and control infrastructure, and the payload`s functionality to exfiltrate confidential data. In order to exfiltrate the data, the attacker sets up single or multiple channels to for data exfiltration on the basis of one or more exfiltration methods as described below. The attacker aims at having the most possible bandwidth with a very low likelihood of being discovered. To achieve this, the attacker usually uses methods which have legitimate uses and are hard to detect among normal network traffic. Following\cite{exfil:1} are the services/features that an attacker abuses to achieve data exfiltration:
\begin{enumerate} 
\item{\textbf{\textit{FTP:}} The attacker sets up an FTP server to collect all the files/data the payload wants to send. The payload running at the target will encapsulate the data in some kind of binary format(possibly compressed and/or encrypted),login to the attacker controlled FTP server and upload the files. Authentication details, ftp server ip addresses are either hard coded or dynamically determined. Such data leak can be detected and stopped by examining packet headers and flow information of FTP traffic. Whitelisting is usually used and network administrators can implement a ceiling on the amount of data any machine in the organization uploads using FTP.}
\item{\textbf{\textit{HTTP POST:}} HTTP POST is a part of HTTP by which data in forms can be sent to the server. The payload can easily perform a HTTP POST request using TCP sockets. Usually file/data is chunked to tackle the inefficiencies of transferring huge data and to behave in line with benign POST requests. The attacker has to setup a web server to accept this POST request and send a response. A typical POST request involves the client requesting an HTML object containing a form from the server following which the client constitutes a POST request with the filled form. In case of data exfiltration there is no initial HTML request. This can be used to detect and prevent HTTP POST exfiltration. A POST request to unknown servers can also be marked suspicious. If not encrypted, POST data can be examined and using regular expressions data leak can be stopped. Attackers nowadays are using SSL to encrypt HTTP POST requests. The attacker implements SSL handshake protocol in the payload, and once the keys are exchanged, the payload sends the data, by encrypting it with the server's public key.
This kind of data leak is the most difficult to detect.  To defend against such breaches, some kind of behavioral system which detects anomalies should be placed at the perimeter.}
\end{enumerate}
The above mentioned two mechanisms are the ones, that are traditionally used by the attackers. Apart from these methods, there exist data exfiltration methods which rely on EMAIL, SSH, Instant Messages, Social Media, etc. These methods don't support huge volume of data but are used to exfiltrate data of strategic nature. Attackers also make use of public file uploading repositories like Dropbox to upload files at some central location which is then retrieved by the attacker. Attackers can also rely on covert channels. The attacker uses optional fields of the DNS, ICMP, IP, TCP, etc protocols to hide control data. The bandwidth offered by these techniques are very low, but it provides high covertness. Detecting such covert\cite{exfil:2} channels require statistical analysis. Statistical profiles of regular traffic are created and any deviation from this regular profile is deemed as malicious. Packet header mangling, deploying DMZ, limiting protocol support and packet regeneration at the perimeter are some of the techniques used to eliminate covert channels.

\subsection{\textbf{Covering Tracks}}
This phase explains the activity of the attacker to cover his tracks, which has been generated in the course of attack. Analysis of this phase may seem to be unnecessary because the attacker has already succeeded in his objectives. But one can gain insightful knowledge of the attacker's tools, techniques and procedures, as well as the weaknesses in the existing security mechanisms, by analyzing this phase. This can be helpful in preparation for countering the future attacks. This phase can be broadly divided into the following parts:
\begin{enumerate}
\item{\textbf{\textit{Data Elimination:}} 
yaIn this technique, the attackers try to delete the data which can help an analyst to do a post-mortem analysis of the attack. This data can include event log files, dropped binaries, etc. The attackers can directly eliminate the data by “safely” deleting it or they can disable the source for generating the data itself. The various methods of achieving this are disabling event logging, modifying system configuration to disable browser history, etc. \cite{phase7:3}}

\item{\textbf{\textit{Data Manipulation:}}
Attackers can manipulate actual facts to frustrate and mislead analysts. This is a more effective technique when compared to data elimination because it is very difficult to detect if done properly. The attackers can manipulate logs and files to obfuscate their identity as well as they can plant evidences which can lead to misattribution. \cite{phase7:2}}

\item{\textbf{\textit{Direct Attacks on Tools and Techniques:}} 
There are some standard tools, which are used for forensics investigation by analysts. These types of attacks are intended to cause a denial of service condition, where a tool will fail to perform its desired function. \cite{phase7:1} Apart from this, attackers can use various anti memory forensics, anti-debugging and anti-VM techniques to hamper dynamic analysis.}
\end{enumerate}

An analyst should not rely totally on the logs obtained from the compromised system because an attacker can always delete or manipulate logs of the system to make the analyst's life tougher. One of the best practices is to send and store logs remotely on a hardened system continuously. This practice may defeat the attacker's intentions of covering his tracks. The usual approach taken by organizations is to look at security logs after the breach is reported. This approach should be avoided, instead log files should be analyzed from network and end-points continuously on a regular basis. This makes sure that an analyst can identify an anomaly in the logs as soon as the attacker tries to delete or manipulate them. An effective way of monitoring the logs is to correlate logs from different sources such as file, communication and process. Another essential approach is to monitor endpoints for breach and respond with distributed forensics and incident response framework such as GRR. \cite{phase7:4}

\section{Summary}
This paper discusses cyber-attacks from an attacker's perspective to help security professionals to have in-depth understanding of \textit{Cyber Attack Thread}. Control Flow of \textit{Cyber Attack Thread} is explained with technical components and corresponding methodologies. Moreover, security  mechanisms against attack thread are discussed with effective implementation using existing solutions. Various techniques for mitigation, prevention and detection at network and system layer are also discussed in detail to help the defender in implementing effective defences.

\end{document}